# Concerns about Modelling of the EDGES Data

Richard Hills, Girish Kulkarni, P. Daniel Meerburg, & Ewald Puchwein

It is predicted[1] that the spectrum of radio emission from the whole sky should show a dip arising from the action of the light from the first stars on the hydrogen atoms in the surrounding gas, which causes the 21-cm line to appear in absorption against the cosmic microwave background. Bowman et al.[2] identified a broad flat-bottomed absorption profile centred at 78 MHz, which could be this feature, although the depth of the profile is much larger than expected. We have examined the modelling process they used and find that their data implies unphysical parameters for the foreground emission and also that their solution is not unique in the sense that we found other simple formulations for the signal that are different in shape but that also fit their data. We argue that this calls into question the interpretation of these data as an unambiguous detection of the cosmological 21-cm absorption signature.

Bowman et al.[2] describe a "physically-motivated" foreground model containing three parameters describing synchrotron emission (magnitude, spectral index and the "running" of the index) and two for ionospheric emission and absorption. They used a linearized version of this model to perform fits with and without the 21-cm feature. Using this model and the data that they released, we obtained essentially identical results, but we note that accommodating the proposed absorption profile requires a change in the foreground model that is much larger than the initial residuals, see Fig. 1a. We also find that the parameters describing the foregrounds have unphysical values, e.g., the parameter associated with brightness temperature of the ionospheric emission is $> 10^4$ K, while that for the astronomical foreground brightness has a large negative value. Full details of the functions fitted, together with the values found for the parameters, are given in the Supplementary Information.

To gain further insight we fitted the full non-linear expression, taking into account the linkage of the emission and absorption by the ionosphere via the temperature of the electrons, $T_e$. The values found for the optical depth of the ionosphere, $\tau_{ion}$, and for $T_e$ are both negative, which is clearly non-physical. We tried restraining these parameters to physically plausible values[3], $\tau_{ion} > 0.005$ at 75 MHz and $200 < T_e < 2000$ K, and we restricted the centre frequency of the absorption profile to lie between 60 and 90 MHz. The results obtained with and without these restrictions on the parameters are shown in Figs. 1b and 1c. Without the restrictions we obtained essentially the same profile as Bowman et al.[2] and the fit is good. (We describe the fit as "good" whenever the rms of the residuals is below 0.03 K.) With the restrictions the fit is poor, the centre of the profile has moved to the upper limit and its depth has increased to about 2 K.

We then explored cases where the ionospheric opacity and temperature are held fixed at reasonable values but higher-order terms are added to the foreground model, using several different formulations. We find that at least five free foreground parameters, in addition to the four absorption profile parameters, are always needed to obtain a good fit. The parameters found for the profile change substantially when different formulations for the foreground are used. Fig. 1d shows an example of this where a good fit was obtained with an amplitude of about 1.1 K for the absorption feature – even larger than that found by Bowman et al[2].

The residuals found when fitting with successively higher numbers of terms in the foreground model are shown in Fig. 2a. Note that at no stage in this process does a distinct absorption-line feature appear. Instead a broad oscillatory feature is present when 2, 3 or 4 terms are used and it is only the addition of the fifth term that produces a large reduction in the residuals. Although it can be argued that higher-order terms are needed to represent the synchrotron foreground accurately, this is not the behaviour expected[4]. In particular, the relatively large value required for the fifth term is not consistent with what is known about the spectrum of the foreground emission in the range 25–400 MHz[5,6,7,8]. Adding higher order foreground terms has simply moved the problem of unphysical parameter values from the ionosphere to the foreground.

It seems possible that the unphysical values are due to residual systematic errors in the data, perhaps arising in the correction for the frequency dependent beam-shape, but if that is the case then it is not clear that model formulations chosen to suit the astronomical foregrounds are the correct way of removing such effects and there is also no clear basis for deciding how many terms should be included. A general concern is that nine parameters are being fitted to data that span ~50 MHz and contain very little real structure with periods shorter than ~10 MHz, so neighbouring data points in the spectrum are strongly correlated. This means that the number of truly independent data points may not be much larger than the number of parameters being fitted.

We next demonstrated that the profile found by Bowman et al.[2] is not a unique solution. The top lines of Figs. 1a and 1b show the residuals after subtracting a five-parameter foreground fit. As already noted, the ionosphere parameters have non-physical values in those cases. The fourth line down in Fig. 2a shows that we obtain very similar residuals by assuming reasonable fixed values for the ionosphere and fitting a five-parameter power law. The residuals do not, however, show an absorption profile but instead show two peaks at around 65 and 90 MHz. Although one can obtain a good fit using the Bowman et al.[2] profile, Fig. 2b, a good fit can also be achieved with two Gaussian emission features of equal height and width, Fig. 2c. With more terms in the foreground model, i.e. the bottom lines in Fig. 2a, the residuals take the form of undulations with a period of ~ 12.5 MHz. We find that a satisfactory fit can then be obtained with just a sine wave, as shown in Fig. 2d. In both these models the total number of free parameters is again nine.

We also found that, with the 12.5 MHz sine wave removed, a good fit was obtained with five foreground parameters and a broad Gaussian absorption profile and that there is then a large covariance between the foreground and signal components. This suggests that these undulations may be what causes the fitting process used by Bowman et al.[2] to produce a profile with a flattened bottom.

Since the proposed 21-cm absorption profile does not match theoretical expectations in either shape or amplitude, it is not clear why it should be preferred to the other forms of signal explored here or to the many more that can be found in the degenerate space between signal and foreground model.

Therefore, although our analysis does not prove that the feature identified by Bowman et al.[2] is absent from their data, we believe the issues that we have raised are such that the evidence for its presence falls well short of the level required to invoke new physics for its explanation.

G.K. acknowledges support from ERC Advanced Grant 320596 'The Emergence of Structure During the Epoch of Reionization'. P.D.M. and E.P. acknowledge support from Senior Kavli Institute Fellowships at the University of Cambridge. P.D.M. also acknowledges support from The Netherlands Organization For Scientific Research (NWO) VIDI grant (dossier 639.042.730).

**Methods**

We use a least square fitting for testing the models presented in the text and the supplementary material. In addition, we derived posterior distributions of model parameters by implementing a likelihood function into the multi-nested sampler Polychord[9,10].

**Data Availability**

For all our analysis we used the 'Data for Figure 1 of Bowman et al. (2018)' in the EDGES Data Release, which is available at http://loco.lab.asu.edu/edges/edges-data-release/.

# Author Information

**Affiliations**

*Cavendish Laboratory, University of Cambridge, Cambridge, UK*
Richard Hills

*Institute of Astronomy, University of Cambridge, Cambridge, UK*
Girish Kulkarni[1], P. Daniel Meerburg & Ewald Puchwein

*Kavli Institute of Cosmology, University of Cambridge, Cambridge, UK*
Girish Kulkarni[1], P. Daniel Meerburg & Ewald Puchwein

*Department of Applied Mathematics and Theoretical Physics, University of Cambridge, Cambridge, UK*
P. Daniel Meerburg

*Kapteyn Astronomical Institute, University of Groningen, Groningen, The Netherlands*
P. Daniel Meerburg

*Van Swinderen Institute for Particle Physics and Gravity, University of Groningen, Groningen, The Netherlands*
P. Daniel Meerburg

**Contributions**

E.P. initiated this study by pointing out the importance of the role of the foreground parameters in the claimed detection of an absorption profile. All the authors participated in the detailed analysis and the writing of the paper.

**Competing Interests**

Declared none.

**Corresponding author**

Correspondence to Richard Hills (richard@mrao.cam.ac.uk)

# Supplementary Information

The detailed derivations, fits and values of the parameters are available at
https://static-content.springer.com/esm/art%3A10.1038%2Fs41586-018-0796-5/MediaObjects/41586_2018_796_MOESM1_ESM.pdf.

---

[1] Present address: Department of Theoretical Physics, Tata Institute of Fundamental Research, Homi Bhabha Road, Mumbai 400005, India

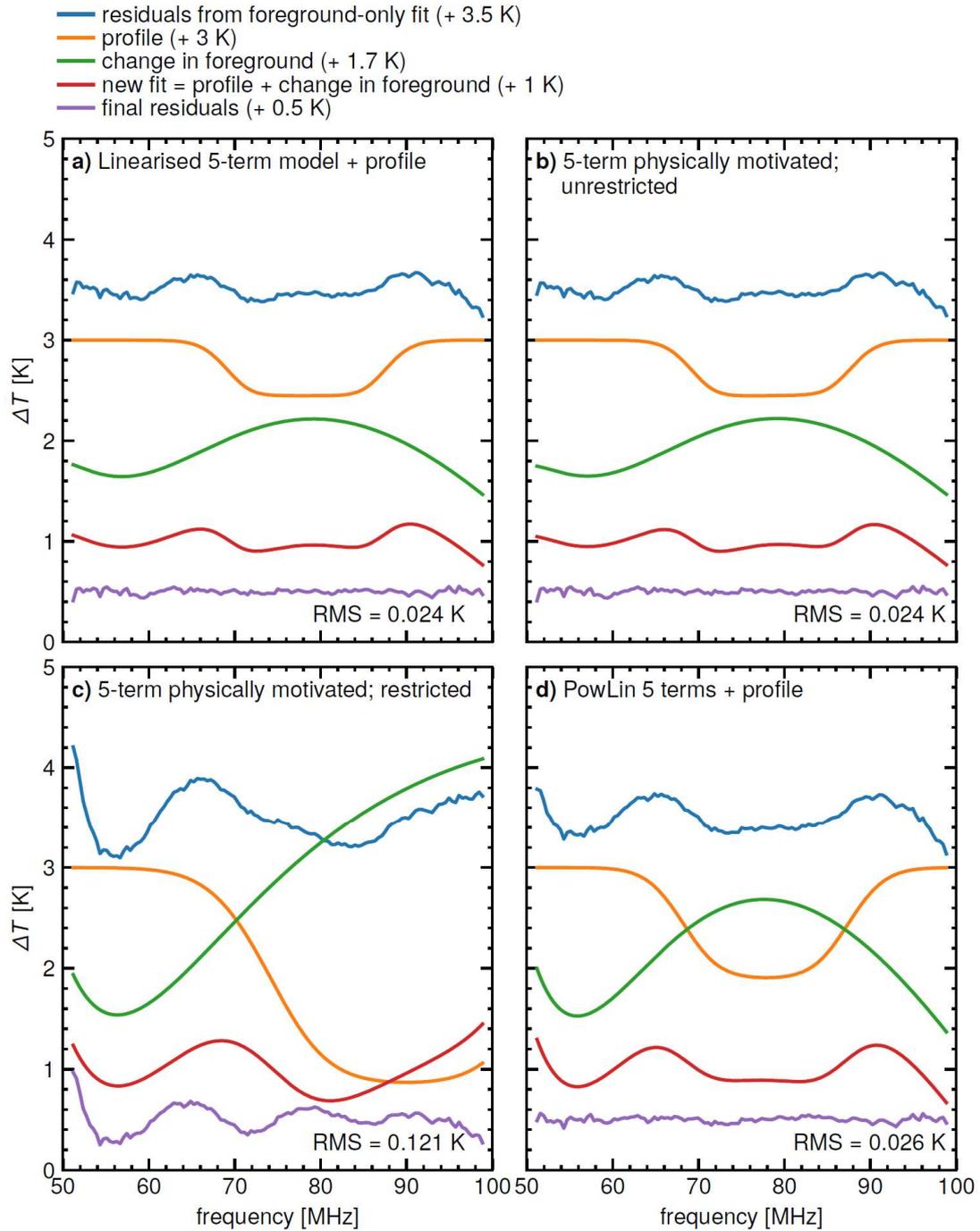

**Fig. 1 | Fits to the EDGES data. a**, With the foregrounds described by the linearized function used by Bowman et al.[2]. **b**, Using the physically motivated nonlinear function for the foregrounds and no restrictions on the parameters. **c**, The same as **b** but with the range of parameters limited to physically plausible values. **d**, Using the PowLin model, which consists of a power law with index given by a polynomial in frequency, $v$. The top line in each panel shows the residuals when a fit is made using the foreground model only. The bottom line is the residual when the fit is run again including the profile with the functional form given by Bowman et al.[2]. The intermediate lines show the shape of the profile found, the change in the foreground model needed to accommodate this and the sum of these two, which is also equal to the difference between the initial and final residuals. The curves have been offset vertically for readability.

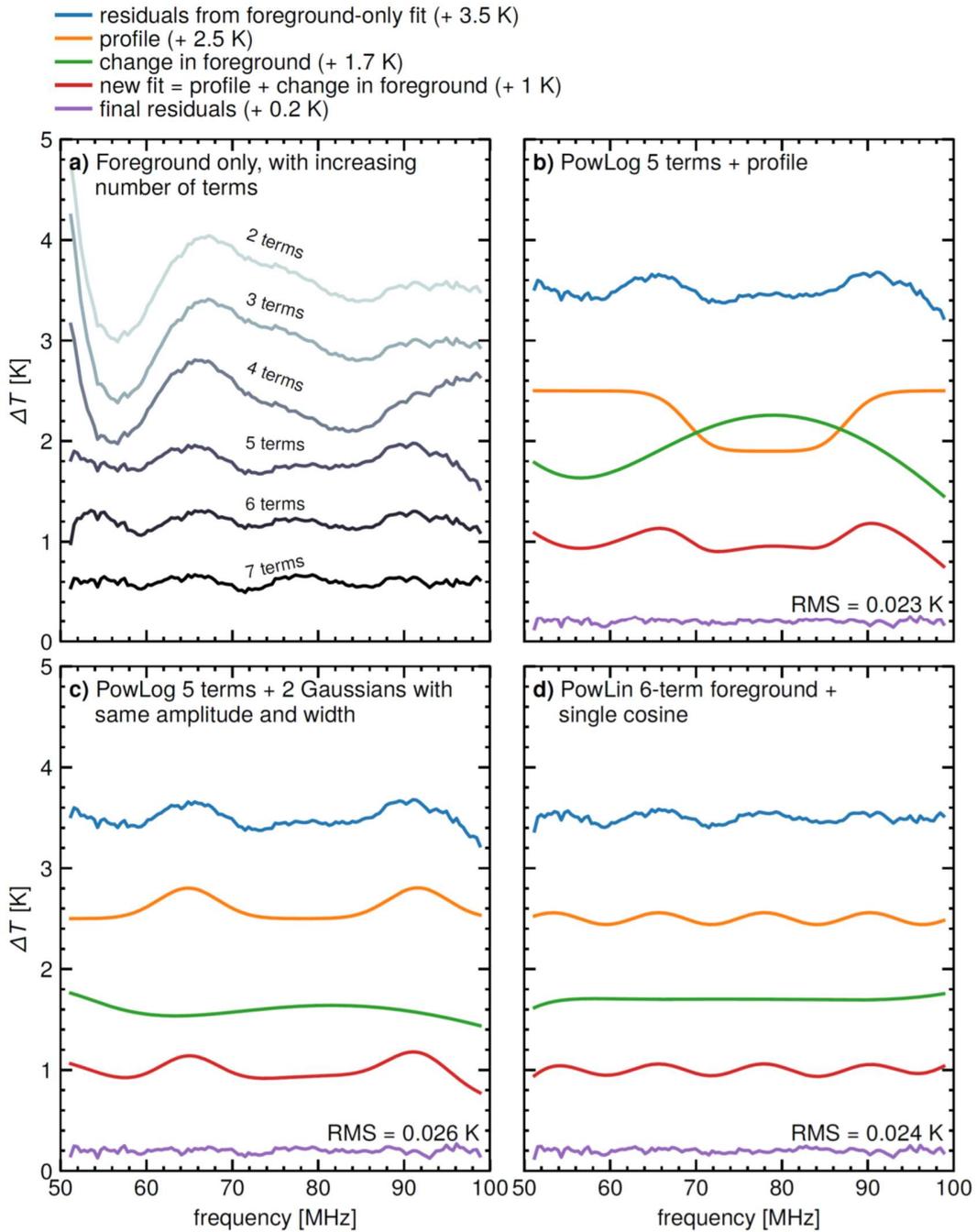

**Fig. 2 | Further illustrations of the fitting process. a**, The residuals when the ionospheric parameters are set to $\tau_0 = 0.014$ and $T_e = 800$ K and the PowLog model – a power law with index given by a polynomial in $\log(v)$ – is fitted with increasing numbers of terms. **b**, Showing how the Bowman et al.[2] profile with four signal parameters can provide a good fit by making use of the freedom provided by five foreground parameters. **c, d**, Alternative nine-parameter fits. **c**, The same model as **b** but with two Gaussian features of equal width and amplitude in place of the Bowman et al.[2] profile. **d**, The PowLin model with six terms and, instead of the profile, a single cosine function, which has a fitted amplitude of about 0.06 K. The different curves have again been offset vertically for readability.